\documentclass{article}
\usepackage{amsmath, amssymb, amsfonts}
\title{Comment on ''Non-vacuum conformally flat space-times: dark energy''} 
\author{Hristu Culetu, \\Ovidius University, Dept.of Physics, \\B-dul Mamaia 124, 900527 Constanta, Romania, \\e-mail : hculetu@yahoo.com}

\begin{document}
\numberwithin{equation}{section}
\pagenumbering{arabic}
\maketitle
\newcommand{\fv}{\boldsymbol{f}}
\newcommand{\tv}{\boldsymbol{t}}
\newcommand{\gv}{\boldsymbol{g}}
\newcommand{\OV}{\boldsymbol{O}}
\newcommand{\wv}{\boldsymbol{w}}
\newcommand{\WV}{\boldsymbol{W}}
\newcommand{\NV}{\boldsymbol{N}}
\newcommand{\hv}{\boldsymbol{h}}
\newcommand{\yv}{\boldsymbol{y}}
\newcommand{\RE}{\textrm{Re}}
\newcommand{\IM}{\textrm{Im}}
\newcommand{\rot}{\textrm{rot}}
\newcommand{\dv}{\boldsymbol{d}}
\newcommand{\grad}{\textrm{grad}}
\newcommand{\Tr}{\textrm{Tr}}
\newcommand{\ua}{\uparrow}
\newcommand{\da}{\downarrow}
\newcommand{\ct}{\textrm{const}}
\newcommand{\xv}{\boldsymbol{x}}
\newcommand{\mv}{\boldsymbol{m}}
\newcommand{\rv}{\boldsymbol{r}}
\newcommand{\kv}{\boldsymbol{k}}
\newcommand{\VE}{\boldsymbol{V}}
\newcommand{\sv}{\boldsymbol{s}}
\newcommand{\RV}{\boldsymbol{R}}
\newcommand{\pv}{\boldsymbol{p}}
\newcommand{\PV}{\boldsymbol{P}}
\newcommand{\EV}{\boldsymbol{E}}
\newcommand{\DV}{\boldsymbol{D}}
\newcommand{\BV}{\boldsymbol{B}}
\newcommand{\HV}{\boldsymbol{H}}
\newcommand{\MV}{\boldsymbol{M}}
\newcommand{\be}{\begin{equation}}
\newcommand{\ee}{\end{equation}}
\newcommand{\ba}{\begin{eqnarray}}
\newcommand{\ea}{\end{eqnarray}}
\newcommand{\bq}{\begin{eqnarray*}}
\newcommand{\eq}{\end{eqnarray*}}
\newcommand{\pa}{\partial}
\newcommand{\f}{\frac}
\newcommand{\FV}{\boldsymbol{F}}
\newcommand{\ve}{\boldsymbol{v}}
\newcommand{\AV}{\boldsymbol{A}}
\newcommand{\jv}{\boldsymbol{j}}
\newcommand{\LV}{\boldsymbol{L}}
\newcommand{\SV}{\boldsymbol{S}}
\newcommand{\av}{\boldsymbol{a}}
\newcommand{\qv}{\boldsymbol{q}}
\newcommand{\QV}{\boldsymbol{Q}}
\newcommand{\ev}{\boldsymbol{e}}
\newcommand{\uv}{\boldsymbol{u}}
\newcommand{\KV}{\boldsymbol{K}}
\newcommand{\ro}{\boldsymbol{\rho}}
\newcommand{\si}{\boldsymbol{\sigma}}
\newcommand{\thv}{\boldsymbol{\theta}}
\newcommand{\bv}{\boldsymbol{b}}
\newcommand{\JV}{\boldsymbol{J}}
\newcommand{\nv}{\boldsymbol{n}}
\newcommand{\lv}{\boldsymbol{l}}
\newcommand{\om}{\boldsymbol{\omega}}
\newcommand{\Om}{\boldsymbol{\Omega}}
\newcommand{\Piv}{\boldsymbol{\Pi}}
\newcommand{\UV}{\boldsymbol{U}}
\newcommand{\iv}{\boldsymbol{i}}
\newcommand{\nuv}{\boldsymbol{\nu}}
\newcommand{\muv}{\boldsymbol{\mu}}
\newcommand{\lm}{\boldsymbol{\lambda}}
\newcommand{\Lm}{\boldsymbol{\Lambda}}
\newcommand{\opsi}{\overline{\psi}}
\renewcommand{\tan}{\textrm{tg}}
\renewcommand{\cot}{\textrm{ctg}}
\renewcommand{\sinh}{\textrm{sh}}
\renewcommand{\cosh}{\textrm{ch}}
\renewcommand{\tanh}{\textrm{th}}
\renewcommand{\coth}{\textrm{cth}}

\begin{abstract}
Ibohal, Ishwarchandra and Singh \cite{IIS} proposed a class of exact, non-vacuum and conformally flat solutions of Einstein's equations whose stress tensor $T_{ab}$ has negative pressure. We show that $T_{ab}$ corresponds to an anisotropic fluid and the equation of state parameter seems not to be $\omega = -1/2$. We consider the authors' constant cannot be the mass of a test particle but is related to a Rindler acceleration of a spherical distribution of uniformly accelerating observers.\\
\end{abstract}

 In their paper \cite{IIS}, Ibohal, Ishwarchandra and Yugindro Singh (IIS) analyzed in detail exact solutions (stationary and non-stationary) of Einstein's equations by especially using the case $n = 2$ in the Wang-Wu mass function expression. We write down the IIS line element of a general metric in Eddington-Finkelstein coordinates $(u, r, \theta, \phi)$
\begin{equation}
 ds^{2} = \left(1 - \frac{2M(u,r)}{r}\right) du^{2} + 2du dr - r^{2} d \Omega^{2},
\label{1}
\end{equation}
 where $M(u,r)$ is the mass function and $u$ is the retarded time coordinate. Using the general form (2.2) of the energy-momentum tensor, they identified the parameters $\mu, \rho$ and $p$ as \cite{IIS}
 \begin{equation}
 \mu = - \frac{2}{Kr^{2}} \frac{\partial M(u,r)}{\partial u},~~~\rho = \frac{2}{Kr^{2}} \frac{\partial M(u,r)}{\partial r},~~~p = - \frac{2}{Kr} \frac{\partial^{2} M(u,r)}{\partial r^{2}}
\label{2}
\end{equation}
with $K = 8\pi G/c^{4}$. One seems to be a typos in the above expression of the pressure $p$ of the fluid in \cite{IIS}, with factor 2 instead of 1. That was rectified in Eq.(2.9) where $p = -\rho/2$.

It is worth to note that the previous expressions for $\mu, \rho$ and $p$ in (0.2) do not depend on the Newtonn's constant $G$. For example, the energy density $\rho$ can be written as (fundamental constants included) 
 \begin{equation}
 \rho = \frac{2c^{4}}{8 \pi Gr^{2}} \frac{G}{c^{2}} \frac{\partial M(u,r)}{\partial r} = \frac{c^{2}}{4 \pi r^{2}} \frac{\partial M(u,r)}{\partial r}
\label{3}
\end{equation}
A similar $G$ - independent expression has been obtained in \cite{HC1} for the heat flux in a time dependent Schwarzschild - de Sitter geometry.

Let us now proceed to the particular case chosen by IIS for the mass function, namely 
 \begin{equation}
 M(u,r) = mr^{2}
\label{4}
\end{equation}
Now the metric (0.1) becomes
\begin{equation}
 ds^{2} = (1 - 2mr)du^{2} + 2du dr - r^{2} d \Omega^{2},
\label{5}
\end{equation}
where $d \Omega^{2}$ is the metric on the unit 2-sphere. The Authors of \cite{IIS} take the constant $m$ as the mass of a test particle present in the spacetime. In our view, the gravitational field generated by a test particle is negligible and its behavior is determined by other (bigger) particle(s) around it. Therefore, the mass $m$ cannot be present in the metric (it cannot curve the spacetime) as is valid for the Schwarzschild line element where the mass is the source of curvature. It is true that $m$ enters the components of the stress tensor but this is in contradiction, in our opinion,  with the role of $m$ as a test particle. We stress that the Schwarzschild geometry is valid in empty space (outside the source, which is introduced 
 by a boundary condition). Moreover, $m$ in (0.5) may not mean a mass for dimensional reasons. In order the product $mr$ to be dimensionless, we must introduce Plank's constant $\hbar$, that is we have to divide $r$ by the Compton wavelength associated to $m$. Why to have $\hbar$ in the metric? The coordinate singularity at $r = 1/2m$, as IIS named the horizon, is actually located at $r = \hbar/2mc$. If $m$ has a microscopic value (of the order of an elementary particle, for instance), $r$ is still too tiny to be of physical interest. 
 
 The situation is worse with the Authors' choice (0.4). Now we must introduce a distance squared to get the same units on both sides of Eq.(0.4). One has
 \begin{equation}
  M(r) = m \left(\frac{r}{l_{P}}\right)^{2},
\label{6}
\end{equation}
where $l_{P} = (G\hbar/c^{3})^{1/2}$ is the Planck length. We again used $\hbar$ to establish the physical dimensions. Since from Eq.(2.1) of \cite{IIS} the relation $r > 2M(u,r)$ should be satisfied, Eq.(0.6) yields $m(r/l_{P})^{2} < r/2$, or $r < \hbar/2mc$, which is in accordance with the fact that $r$ should be less than the horizon distance. The expression (0.3) for the energy density is now given by
 \begin{equation}
 \rho = \frac{1}{2 \pi l_{P}^{2}} \frac{mc^{2}}{r}.
\label{7}
\end{equation}
Compared to (0.3), now $\rho$ depends on $\hbar$ that is located at the denominator. It seems to be extremely large to be viable. Apart from its divergenve at $r = 0$ (a true singularity as IIS have correctly noted), $\rho$ is huge even for a small ratio $m/r$. With $m \approx 10^{-24} g$ (the proton mass) and $r = R_{U} = 10^{28} cm$, we get $\rho \approx 10^{36} erg/cm^{3}$. As $m$ from (2.7) of \cite{IIS} appears to be a source mass, it is unconceivable to have the above value of $\rho$ at a distance from the singularity of the order of the Universe's radius $R_{U}$. Let us take now for $m$ the minimal value $10^{-65}$g from Eq. (2.35) of \cite{IIS}. With $r$, say, of $10^{5}$ cm, we obtain from (0.6) that $M = m (r/l_{P})^{2} \approx 10^{11}g$. That means the mass comprised in a sphere with radius $1 Km$ is $10^{5}$ tones ($m$ represents the mass comprised in a sphere of radius $l_{P}$ as can be easily seen from (0.6)). As far as the energy density (0.7) is concerned, we have $\rho \approx 10^{16} g/cm^{3}$ at $r = 1 Km$ from the origin of coordinates. We consider the above values of $M$ and $\rho$ as being too large to be realistic, even though one is dealing with dark energy here.

We therefore reach the conclusion that the constant $m$ in (2.7) of \cite{IIS} cannot be a mass. Moreover, the Authors of \cite{IIS} did not take into account any quantum field in the game and therefore $m$ may not be considered as the particle of the field. So we call on the papers \cite{DG, HC2, HC3} and consider $m$ to be a Rindler acceleration (noting that we were aware of \cite{IIS} once the version arXiv: 0903.3134v4 was issued (July 1, 2011)). Now the metric coefficient $1 - 2mr$ becomes $1 - 2mr/c^{2}$, with no use of $\hbar$. The horizon will be located at $r = c^{2}/2m$, that is macroscopic for reasonable values of the constant acceleration $m$. The spacetimes (2.38) or (2.7) from \cite{IIS} correspond exactly to the geometry (2.1) from \cite{HC3} or Eq.(16) from \cite{HC2}. It is worth to remind that Mannheim \cite{PM} reached a Schwarzschild solution with a Rindler term for a generalization of Einstein's equations starting with a conformally-invariant Lagrangean (the Weyl tensor squared). 

A similar calculus for the energy density $\rho$ could be performed with $m$ an acceleration. Now taking, say, $m = 10^{3} cm/s^{2}$ and $r = R_{U}$, one obtains from Eq.(2) of \cite{HC2} (with $m$ instead of $g$) that $\rho = c^{2}m/2\pi GR_{U} \approx 1 ~erg/cm^{3}$ which turns out to be much more reasonable than $\rho \approx 10^{36}$ obtained before.

IIS actually anticipated our interpretation of the constant $m$ when they stated that ''...the surface gravity is directly measured by the mass or, in other words, it is directly proportional to the mass of the solution''(below Eq.(2.33), p.9). 

We now show that the anisotropic fluid stress tensor (2) from \cite{HC2} is exactly the energy momentum tensor (2.9) from \cite{IIS}. We note firstly that $p$ from (2.3) of \cite{IIS} is  the trasversal pressure $p_{\bot}~ (p_{\theta}$ or $p_{\phi}$) because it is the coefficient of the term containing $m_{a} \bar{m}_{b} + \bar{m}_{a}m_{b}$, which has only angular nonzero components. That is easier seen from the following relations
  \begin{equation}
  \rho = T^{u}_{~u},~~~p_{r} = -T^{r}_{~r} = -\rho,~~~p_{\bot} = -T^{\theta}_{~\theta} = -T^{\phi}_{~\phi} = \frac{p_{r}}{2},
\label{8}
\end{equation}
where $T^{a}_{~b}$ are the components of (2.8) from \cite{IIS} and $p_{r}$ is the radial pressure (noting that $\mu = T^{r}_{~u}$ is related to the heat flux due to the nonstationary character of the metric (2.1)). IIS observed the fluid (2.2) is not perfect (even when $\mu = 0$) but they did not emphasized that it is anisotropic, with $p_{r} = p_{\bot}/2$. They kept track only of $p$, not specifying that it is actually a tangential pressure. IIS did not comment their Eqs.(2.13) to clarify the role of $p_{r} = - \rho$ and to find that in fact $\omega = p_{r}/\rho = -1$ and not $\omega = p_{\bot}/\rho = -1/2$.

The fact that the two forms of the anisotropic stress tensor (Eq. (3.4) from \cite{IIS} with $\mu = 0$ and Eq. (5) from \cite{HC2}) are identical can be easily seen if we take into account that $s_{a}$ from (5) plays the role of $v_{a}$ from (3.4), $p_{r}$ is $(-\rho)$ and $p$ is $p_{\bot}$ (noting that in \cite{HC2} the positive signature is used). 

Another observation concerns the value of the Kretschmann scalar. Using the software package Maple - GRTensor, we found for the spacetime (2.7) from \cite{IIS} the expression $32m^{2}/r^{2}$ and not $-160m^{2}/r^{2}$, as IIS obtained in their Eq. (2.24). There is, perhaps, a typos here.

The Authors of \cite{IIS} have found that all tetrad components of the Weyl tensor vanish, i.e. their metric (2.38) is conformally-flat. However, they have not written down its conformally-flat version. We note, as an aside, that that version has been obtained, to the best of our knowledge, 10 years ago by Kiselev \cite{VVK} (see also \cite{HC4}) and reads
 \begin{equation}
 ds^{2} = \frac{1}{m^{2}(T + R)^{2}} (dT^{2} - dR^{2} - R^{2} d\Omega^{2})
\label{9}
\end{equation}  
 with $T = e^{mt}cosh \chi,~R = e^{mt}sinh \chi$ and $1 - 2mr = e^{-2\chi}$ ($\chi > 0,~T > 0, ~0 <R <T)$. 
 
 Let us now search the expression (2.31) from \cite{IIS} for the horizon entropy. Whilst the entropy of a black hole is proportional to its mass squared, IIS obtained, using the same formula $S = A/4$ ($A$ is the horizon area) that $S \propto 1/m^{2}$. In both spacetimes (Schwarzschild's and, respectively, (2.7)), $m$ is the source of curvature. An entropy that diminishes when the source mass increases seems to be against the common sense. 
 
 Why did IIS consider $r > 10^{28} cm \equiv R_{U}$ above their Eq.(2.35)? $R_{U}$ is the radius of the Universe (cosmological horizon) and we cannot have $r > R_{U}$ (consider, say, the static de Sitter universe). Taking $r > R_{U}$ is equivalent with a shift of $R_{U}$ that leads to a new (larger) cosmological horizon. Moreover, above Eq.(2.35) IIS localize the mass $m \approx 10^{-60}$ or $m \approx 10^{-65}g$ at the horizon $r = 1/2m$. To have a mass of that magnitude localized on a sphere of radius $10^{28} cm$ does not seem to have, classically, any physical meaning. In contrast, taking $m$ an acceleration, one has $m = c^{2}/2R_{U} \approx 10^{-8} cm/s^{2}$ that equals the universal critical acceleration from MOND paradigm \cite{BCCL, HC2}. In addition, one is not allowed to compare the magnitude of $m$ with that of the cosmological constant $\Lambda$ (as IIS did below Eq.(2.35)) because they have different dimensions.

\end{document}